\documentstyle[12pt,epsfig,rotating]{article}

\newcommand{\be}{\begin{eqnarray}}
\newcommand{\ee}{\end{eqnarray}}
\newcommand{\dia}{\!\!\!\!\!\!\not\,\,\,\,\,}

\title{Damping and reaction rates and
       wave function renormalization of
       fermions in hot gauge theories}
\author{ {\bf Alejandro Ayala}, {\bf Juan Carlos D'Olivo}\\
         {\it Instituto de Ciencias Nucleares}\\
         {\it Universidad Nacional Aut\'onoma de M\'exico}\\ \vspace*{0.2in}
         {\it Aptartado Postal 70-543, M\'exico Distrito Federal 04510,
         M\'exico.}\\
         {\bf Axel Weber}\\
         {\it Instituto de F\'{\i}sica y Matem\'aticas}\\
         {\it Universidad Michoacana de San Nicol\'as de Hidalgo}\\
         {\it Apartado Postal 2-82 Morelia Michoac\'an 58040, M\'exico.} }

\begin{document}

\maketitle
\begin{abstract}

We examine the relation between the damping rate of a chiral fermion mode 
propagating in a hot plasma and the rate at which the
mode approaches equilibrium. We show how these two quantities, obtained
from the imaginary part of the fermion self-energy, are equal when the
reaction rate is defined using the appropriate wave function of the mode in 
the medium. As an application, we compute the production rate of
hard axions by Compton-like scattering processes in a hot QED plasma 
starting from both, the axion self-energy and the electron 
self-energy. We show that the latter rate coincides with the former only
when this is computed using the corresponding medium spinor modes.

\end{abstract}

\begin{flushleft}
\end{flushleft}

\newpage

\section{Introduction}

As is well known, the properties of a chiral fermion propagating in a 
medium are modified from those in vacuum. Interactions give rise to the 
appearance of collective modes characterized by dispersion relations and 
damping rates. To be precise, if $P^{\mu}=(\omega,\vec{p})$ is the 
four-momentum of a fermion mode in the rest frame of the medium, the 
corresponding pole of the propagator is located, in general, 
at a complex value
\be
   \omega=\omega_p-i\,\frac{\gamma}{2}\, ,
   \label{eq:pole}
\ee
where $\omega_p$ and $\gamma$ are real functions of $p$ and correspond to 
the energy and damping rate of the mode, respectively. In particular, 
$m_f\equiv \omega_{p=0}$ 
can be interpreted as the mass of the fermionic excitation and 
$\gamma^{-1}$, in accordance with linear response theory~\cite{LeBellac}, 
gives the characteristic time scale for the decay of the mode. In a field 
theoretical description, $\omega_p$ and $\gamma$ are determined (for
weak damping) from the real and imaginary part of the fermion self-energy, 
respectively.

We recall, however, that there exists a distinct concept also related to
the imaginary part of the fermion self-energy. This quantity, which we will
call the total reaction rate $\Gamma$, is interpreted in such a way
that its inverse gives the time scale for the fermion 
distribution to approach equilibrium~\cite{Weldon1,Kadanoff}.
It would then appear that one has two different physical quantities, 
the damping rate $\gamma$ and the total reaction rate
$\Gamma$, both obtained from the imaginary part of the fermion self-energy.
This has indeed been a source of some confusion in the literature, for
instance, in a paper on the subject, two different expressions are
used to compute the damping and total reaction rate in terms of the 
self-energy~\cite{Altherr}.

The equivalence between the above two quantities was established by D'Olivo 
and Nieves~\cite{D'Olivo} for a chiral fermion interacting with a scalar and 
a massive fermion through a Yukawa coupling. They showed that in this case
\be
   \Gamma=\gamma\, ,
   \label{eq:equiv}
\ee
provided $\Gamma$ is defined using the appropriate wave functions of the 
chiral fermion mode in the medium. In this paper we show explicitly how
this equivalence is realized in the realm of hot gauge theories where the 
resummation method of Braaten and Pisarski~\cite{Braaten} has provided a 
consistent framework for the computation of the leading-order contributions to
damping rates. We also stress the role played by wave function renormalization
of the medium spinor modes for the above equivalence. We illustrate this
role by computing the production rate of hard axions in a hot QED plasma 
through Compton-like scattering processes, showing that the rate obtained
by considering the axion self-energy coincides with the rate obtained
from the expression for the electron self-energy only when the appropriate 
medium spinors are used. 

This paper is organized as follows: Sections II and III are devoted to a 
brief recollection of some well established concepts regarding the 
self-energy of a chiral fermion in a medium and its relation to the fermion
damping rate. In section II we review how the damping rate can be written in 
terms of the imaginary part of the self-energy and the spinors satisfying the 
effective Dirac equation in the medium. In section III we briefly recall how a
complete leading order calculation yields gauge independent poles for the 
fermion propagator in the medium in a covariant gauge. In section IV we 
express the imaginary part of the self-energy as a total reaction rate. We 
identify the physical processes involved and establish the equivalence between
the damping rate and the total reaction rate when the latter is expressed in 
terms of the spinors in the medium. In section V we compute the production 
rate of hard axions in a hot QED plasma. We finally discuss our results in 
section VI. A short appendix collects notations and definitions of some 
quantities appearing in section IV.

\section{The damping rate}

In this section we express the damping rate in terms of the self-energy, 
following D'Olivo and Nieves~\cite{D'Olivo}, thereby establishing our notation 
and preparing for the discussion in the next section. The inverse propagator 
of a chiral fermion in a medium can be written in 
general as
\be
   iS^{-1}(P)=P\dia - \,\Sigma\, ,
   \label{eq:propmed}
\ee
where $\Sigma$ is the effective self-energy induced by the medium.
Hereafter, capital letters are used to refer to momentum four-vectors.
At $T=0$ the theory is chirally invariant, due to the vanishing of the 
fermion mass. Chiral invariance is unaffected by the presence of the medium, 
hence the inverse propagator in the medium rest frame can be expressed as
\be
   iS^{-1}(P) &=& A_0\gamma_0-A_s\vec{\gamma}\cdot\hat{p}\, 
   \nonumber \\
   &=& \frac{1}{2}\,\Delta_+^{-1}(P)\,(\gamma_0 +\vec{\gamma}\cdot\hat{p}) +
   \frac{1}{2}\,\Delta_-^{-1}(P)\,(\gamma_0 -\vec{\gamma}\cdot\hat{p})
   \label{eq:moregenprop}
\ee
in terms of the functions $\Delta_{\pm}(P)=(A_0\mp A_s)^{-1}$,
where we have denoted the unit vector $\vec{p}/p$ by $\hat{p}$.
The propagator poles are given by $\Delta_{\pm}^{-1}(P)=0$. In the 
positive-energy sector, Eq.~(\ref{eq:moregenprop}) yields four propagating 
modes, two with positive and two with negative helicity over chirality ratio 
\cite{Weldon2}. For definitiveness, let us work with a 
negative-helicity solution and look for the positive-energy poles in
$\Delta_+(P)$. The same arguments apply to negative-energy and 
positive-helicity solutions and the poles of $\Delta_-(P)$. 

$\Delta^{-1}_+(\omega,p)$ has in general a real and an imaginary part. 
Writing $\omega$ as in Eq.~(\ref{eq:pole}), the equation determining the
poles is
\be
   \mbox{Re}\,\Delta^{-1}_+(\omega_p-i\gamma /2,p) +
   i\, \mbox{Im}\,\Delta^{-1}_+(\omega_p-i\gamma /2,p) 
   =0\, .
   \label{eq:poleeq}
\ee
In the following, we consider the case of weak damping, 
$\gamma\ll\omega_p$, where the physical concept of a propagating mode is
still meaningful. Eq.~(\ref{eq:poleeq}) can then be solved by 
expanding it in powers of $\gamma$ and retaining terms at most linear in 
$\gamma$ and $\mbox{Im}\,\Delta^{-1}_+$. At zeroth order, $\omega_p$ is 
obtained from
\be
   \mbox{Re}\,\Delta^{-1}_+(\omega_p,p)=0\, ,
   \label{eq:omesol}
\ee
whereas $\gamma$ is given by the first-order equation, which yields
\be
   \frac{\gamma}{2}=Z_p\, \mbox{Im}\,\Delta^{-1}_+(\omega_p,p)\, ,
   \label{eq:gamsol}
\ee 
where 
\be
   Z^{-1}_p=\left[\frac{\partial}{\partial\omega}\,\mbox{Re}
   \,\Delta^{-1}_+\right]_{\omega=\omega_p}\, .
   \label{eq:norm}
\ee
The factor $Z_p$ coincides with the residue at the pole of the one-particle
contribution to the propagator and it is thus the normalization factor that
has to be taken into account to construct the spinors representing 
one-particle states~\cite{Nieves}. In the context of many-body physics,
$Z_p$ represents the probability factor that needs to be considered for the 
various processes involving the fermion mode~\cite{Kirzhnits}.

In order to relate $\gamma$ to $\mbox{Im}\,\Sigma$, recall
that the one-particle states are represented by spinors satisfying the
effective Dirac equation
\be
   (P\dia\! - \,\mbox{Re}\,\Sigma)u(P)=0
   \label{eq:eff1}
\ee
obtained by neglecting the absorptive part of the effective self-energy.
Real and imaginary part of $\Sigma$ are defined, as usual, by
\be
   \mbox{Re}\,\Sigma &=& \frac{1}{2} \left(\Sigma + 
   \gamma^0\Sigma^\dagger\gamma^0 \right) \, , \nonumber \\
   \mbox{Im}\,\Sigma &=& \frac{1}{2i} \left(\Sigma - 
   \gamma^0\Sigma^\dagger\gamma^0 \right) \, .
\ee
In the following, we will not need the explicit solutions of 
Eq.~(\ref{eq:eff1}), but rather work with the corresponding projection 
operators. Choosing a normalization corresponding to that 
of the one particle states~\cite{Nieves}, the projector for the solution 
under consideration can be written as
\be
   u(P)\bar{u}(P)=Z_p\,\omega_p\, L\, n\dia\, ,
   \label{eq:proj}
\ee
where $L=(1 - \gamma_5)/2$ and $n^{\mu}$ is a four-vector
with components $n^{\mu}=(1,\hat{p})$ in the plasma rest frame. From 
Eqs.~(\ref{eq:propmed}), (\ref{eq:moregenprop}) and (\ref{eq:gamsol}) we get, 
by a straightforward computation,
\be
   \gamma&=&-\frac{Z_p}{2}\,\mbox{Tr}\,[n\dia
   \!\mbox{Im}\,\Sigma(\omega_p,p)]\nonumber\\
   &=&-\frac{1}{\omega_p}\,\bar{u}(P)\,
   \mbox{Im}\,\Sigma(\omega_p,p)\,u(P)\, ,
   \label{eq:some2}
\ee
where in the last line we have used Eq.~(\ref{eq:proj}). Note that 
in Eq.~(\ref{eq:some2}) the self-energy is evaluated at $\omega=\omega_p$.
Eq.~(\ref{eq:some2}) is valid independently of any approximation scheme,
perturbative or not, and it only requires that the damping rate be small 
compared to the energy of the mode. This equation was already used in 
Ref.~\cite{D'Olivo} to establish the equality of damping and reaction rates in
a model with Yukawa coupling. In the next section, we review how the damping 
rate is obtained in hot gauge theories.

\section{The HTL self-energy}

In a gauge theory, a plasma of massless particles is characterized by two
quantities, the temperature $T$ and the dimensionless coupling constant
$g$. If $g\ll 1$ there exist two distinct energy scales, $T$ and $gT$. The 
first is characteristic of individual particles, whereas the second
is typical of collective phenomena. In the Hard Thermal Loop (HTL) 
approximation, bare perturbation theory is used to compute the leading 
contributions to the self-energies and vertices. For soft external 
momenta (i.e.~of order $\sim gT$), these are of the same order of magnitude 
as the bare quantities that they modify, thus the necessity of resummation.
HTLs provide effective propagators and vertices necessary for a consistent
perturbative expansion. In the following, we will consider exclusively
soft external momenta. The lowest order HTL contribution to the fermion
propagator can be shown to be real (for time-like momenta) and 
gauge-independent~\cite{Braaten}. A complete leading order calculation of 
the imaginary part of the self-energy requires, however,
the sum of the two diagrams depicted in Fig.~1. The heavy dots
indicate effective propagators and vertices. Fig.~1a is the
usual diagram for the fermion self-energy modified by the use of effective
propagators and the effective two-fermion one-gauge boson vertex. 
Fig.~1b is an additional contribution coming from the effective 
two-fermion two-gauge boson vertex, upon contracting the gauge boson lines. 

\begin{figure}[b]
\vspace{.2in}
\centerline{\epsfig{file=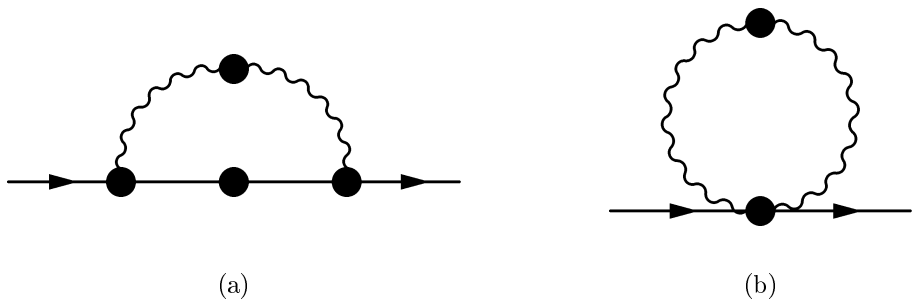}}
\vspace{.2in}
\caption{ The self-energy graphs for fermions at one-loop level in the 
          effective expansion. Heavy dots denote the effective propagators and 
          vertices arising in the HTL approximation }
\end{figure}

Let us denote by $^\star\Sigma(P)$ the effective self-energy calculated
to leading order in its real and imaginary parts. In the imaginary
time formalism of Thermal Field Theory, the contributions from Figs.~1a 
and 1b are, respectively
\be
   \,^\star\Sigma_a(P)&=&-g^2C_FT\sum_n\int\frac{d^3k}{(2\pi)^3}
   \Gamma_{\mu}(P,P-K) D_{\mu\nu}(K) S(P-K) \Gamma_{\nu}(P-K,P)
   \nonumber\\
   \,^\star\Sigma_b(P)&=&-\frac{g^2}{2}C_FT\sum_n\int\frac{d^3k}{(2\pi)^3}
   \Gamma_{\mu\nu}(P,P,K) D_{\mu\nu}(K)\, ,
   \label{eq:sigma}
\ee
where $C_F$ is the Casimir invariant corresponding to the fundamental
representation of the gauge group. The effective two-fermion one-gauge boson 
and two-fermion two-gauge boson vertices are denoted by 
$\Gamma_{\mu}$ and $\Gamma_{\mu\nu}$, respectively. 
$D_{\mu\nu}$ and $S$ represent the gauge boson and fermion
effective propagators, respectively. 

In a covariant gauge, $D_{\mu\nu}$ has a piece proportional to the gauge
parameter $\xi$. Let us separate its contribution to the effective self-energy
writing
\be
   ^\star\Sigma(P)=\,^\star\Sigma_0(P) +\,^\star\Sigma_\xi(P)
   \label{eq:normterm}
\ee
where $^\star\Sigma_0(P)$ is gauge-parameter independent and 
$^\star\Sigma_\xi(P)$ is the contribution to Eq.~(\ref{eq:sigma}) from the 
term proportional to the gauge parameter in $D_{\mu\nu}$, namely, from
\be
   \frac{\xi}{K^2}\frac{K_{\mu}K_{\nu}}{K^2}\, .
   \nonumber
\ee
By using the Ward identities satisfied by the HTLs we can express the
gauge-parameter dependent term in Eq.~(\ref{eq:normterm}) as
\be
   ^\star\Sigma_\xi(P)=a(P)S^{-1}(P)\, ,
   \label{eq:normexpl}
\ee
where we have defined
\be
   a(P)\equiv\xi g^2C_F T\sum_n\int\frac{d^3k}{(2\pi)^3}
   \left[\frac{1}{(K^2)^2}-\frac{S(P-K)S^{-1}(P)}{(K^2)^2}\right]\, .
   \label{eq:apar}
\ee
The function $a(P)$ is infrared singular. In what follows we assume that
a proper infrared regulator has been introduced and that this will be taken
to zero only until the very end of any calculation~\cite{Rebhan}.

The modification that $^\star\Sigma(P)$ introduces in the fermion propagator 
can be written, after analytical continuation to Minkowski space, as in
Eq.~(\ref{eq:propmed}). Using Eqs.~(\ref{eq:normterm}) and (\ref{eq:normexpl}),
we have
\be
   i\,^\star S^{-1}(P)=P\dia\!
   - \,^\star\Sigma_0(P) - i\,a(P)S^{-1}(P)\, .
   \label{eq:modprop}
\ee
A simple power counting shows that $a(P)$ is of order $g$. 
Therefore, to the given order, it is allowed to replace $S^{-1}(P)$ by
$^\star S^{-1}(P)$ in Eq.~(\ref{eq:modprop}) which now becomes
\be
   i\,^\star S^{-1}(P)&=&P\dia\! - \,^\star\Sigma(P)\nonumber\\
   &\simeq&[1-a(P)]
   [P\dia\! - \,^\star\Sigma_0(P)]
   \, ,
   \label{eq:replace}
\ee
whence 
\be
   ^\star\Sigma(P)\equiv a(P)P\dia\! + \, [1-a(P)]
   \,^\star\Sigma_0(P) \, .
   \label{eq:replace1}
\ee
Notice that in principle, there are other non-leading contributions to the real 
part of $^\star\Sigma(P)$ which are not taken into account in 
Eq.~(\ref{eq:replace1}). In fact, for the calculation of the damping rate
to leading order, it is only necessary to consider the leading contributions
to the real part of $^\star\Sigma(P)$. Nevertheless, in the following we keep
a non-leading term containing $a(P)$ to illustrate that its presence does
not affect the gauge independence of the damping rate.

Eq.~(\ref{eq:replace}) implies that the poles of the propagator are 
gauge-parameter independent, since a pole of 
$[P\!\dia\! - \,^\star\Sigma_0(P)]^{-1}$ is
also a pole of $^\star S$. Within our approximation, the 
effective Dirac equation for the one-particle fermion mode is
\be
   [P\dia\! - \, {\mbox R}{\mbox e}\,^\star\Sigma(P)]
   \,^\star u(P)=0\, .
   \label{eq:effective}
\ee
By neglecting the absorptive with respect to the dispersive part of 
$^\star\Sigma_0(P)$, Eq.~(\ref{eq:effective}) becomes
\be
   [P\dia\! - \,
   {\mbox R}{\mbox e}\,^\star\Sigma_0(P)]\,^\star u(P)=0\, .
   \label{eq:realeffec}
\ee
According to Eq.~(\ref{eq:proj}), a factor 
$[1-{\mbox R}{\mbox e}\, a(P)]^{-1/2}$ has to be absorbed into the 
normalization of the spinor $^\star u(P)$ since this factor is part of the 
residue of the fermion propagator at the pole. 

From. Eqs.~(\ref{eq:some2}) and (\ref{eq:replace1}), the damping rate 
$\gamma$ can be written as
\be
   \gamma&=&-\frac{1}{\omega_p}\,^\star\bar{u}(P)\,
   {\mbox I}{\mbox m}\,^\star\Sigma(\omega_p,p)\,^\star u(P)\nonumber\\
   &=&-\frac{1}{\omega_p}\,[1-{\mbox R}{\mbox e}\,a(P)] 
   \,^\star\bar{u}(P)\,{\mbox I}{\mbox m}\,^\star\Sigma_0(\omega_p,p)
   \,^\star u(P)\, ,
   \label{eq:imsig}
\ee
where in the last line, we have used the effective Dirac equation, 
Eq.~(\ref{eq:realeffec}), satisfied by $^\star u(P)$. $\omega_p$ is the
dispersion relation of the corresponding mode and is given, for example in
Ref.~\cite{LeBellac2}. Let us now define the spinor $u(P)$ related to 
$^\star u(P)$ by
\be
   u(P)\equiv[1-{\mbox R}{\mbox e}\,a(P)]^{1/2}\:^\star u(P)\, ,
   \label{eq:newesp}
\ee
thereby cancelling the gauge-parameter dependent factor in the normalization
of $^\star u(P)$. The spinor $u(P)$ satisfies Eq.~(\ref{eq:realeffec}) and is 
normalized according to Eq.~(\ref{eq:proj}). The explicit expression for the 
residues of the one particle contributions to the propagator 
is~\cite{LeBellac2}
\be
   Z_p&=&\frac{\omega_p-p^2}{2m_f^2}\, .
   \label{eq:res}
\ee
Eq.~(\ref{eq:imsig}) is now written in terms of $u(P)$ as
\be
   \gamma&=&-\frac{1}{\omega_p}\,
   \bar{u}(P)\,{\mbox I}{\mbox m}\,^\star\Sigma_0(\omega_p,p)\,u(P)\, ,
   \label{eq:newesp2}
\ee
which is manifestly gauge-parameter independent. Arguments to show that the 
expression for the damping rate in the Coulomb gauge coincides with that in a 
covariant gauge have been given elsewhere~\cite{Braaten}, therefore all of the 
above translates as well to the case of the Coulomb gauge. In the next section,
we will write the explicit expression for the imaginary part of the
gauge independent piece of the self-energy, $^\star\Sigma_0$.

\section{The reaction rate}

We now proceed to establish the relation between $^\star\Sigma_0$ and
the total reaction rate $\Gamma$ in the HTL approximation. In order to do this,
we need an explicit expression for the imaginary part of $^\star\Sigma_0$. The
imaginary-time expression for the gauge-independent part of the self-energy
is obtained from Eq.~(\ref{eq:sigma}). To evaluate the sum, we use the 
identity (see for example Ref.~\cite{Braaten2})
\be
   &{\mbox I}{\mbox m}&\, T \, 
   \sum_n g(i\omega_n)\tilde{g}(i(\omega-\omega_n))=\pi(e^{p_0/T}+1)
   \nonumber \\
   &\times& \int_{-\infty}^{\infty}\frac{dk_0}{2\pi}
   \int_{-\infty}^{\infty}\frac{dp_0'}{2\pi}
   f(k_0)\tilde{f}(p_0')\delta(p_0-k_0-p_0')\rho(k_0)\tilde{\rho}(p_0')\, ,
   \label{eq:disc}
\ee
where the analytical continuation $i\omega\rightarrow p_0+i\epsilon$ has been
performed. Here $f$ and $\tilde{f}$ represent the statistical distributions, 
whereas $\rho$ and $\tilde{\rho}$ are the spectral densities corresponding to 
the functions $g$ and $\tilde{g}$, respectively. The spectral densities contain
the discontinuities across the real energy axis. Their support depends on 
whether the momentum four-vector is inside or outside the light-cone. 
For time-like momenta, these spectral densities
have support on the quasi-particle poles, whereas for space-like momenta
they have support on the whole interval corresponding to the branch cut of
the propagator. 

Adding up the two contributions in Eq.~(\ref{eq:sigma}), 
${\mbox I}{\mbox m}\,^\star\Sigma_0(\omega_p,p)$ can be 
expressed, by means of Eq.~(\ref{eq:disc}), as an integral involving products 
of spectral densities~\cite{explain},
\be
   {\mbox I}{\mbox m}\,^\star\Sigma_0(\omega_p,p)
   &=&\frac{\alpha_g}{2}C_F(e^{\omega_p/T}\!+1)
   \int\frac{d^3k}{(2\pi)^3}\int_{-\infty}^{\infty} dk_0 dp_0'
   \delta(\omega_p-k_0-p_0')f(k_0)\tilde{f}(p_0')\nonumber\\
   &\times&\left\{
   \gamma^{\mu}\rho_{\mu\nu}(k_0,k)\tilde{\rho}(p_0',\vec{p}-\vec{k})
   \gamma^{\nu}\right.\nonumber\\
   &&+\,\gamma^{\mu}\rho_{\mu\nu}(k_0,k)
   \tilde{\rho}(p_0',\vec{p}-\vec{k})F_1^{\nu}(P,K)\nonumber\\
   &&+\,\gamma^{\mu}\rho_{\mu\nu}(k_0,k)
   \tilde{\alpha}(p_0',\vec{p}-\vec{k})F_2^{\nu}(P,K)\nonumber\\
   &&+\,F_1^{\mu}(P,K)\rho_{\mu\nu}(k_0,k)
   \tilde{\rho}(p_0',\vec{p}-\vec{k})\gamma^{\nu}\nonumber\\
   &&+\,F_2^{\mu}(P,K)\rho_{\mu\nu}(k_0,k)
   \tilde{\alpha}(p_0',\vec{p}-\vec{k})\gamma^{\nu}\nonumber\\ 
   &&+\,F_1^{\mu}(P,K)\rho_{\mu\nu}(k_0,k)
   \tilde{\rho}(p_0',\vec{p}-\vec{k})F_1^{\nu}(P,K)\nonumber\\
   &&+\,F_1^{\mu}(P,K)\rho_{\mu\nu}(k_0,k)
   \tilde{\alpha}(p_0',\vec{p}-\vec{k})F_2^{\nu}(P,K)\nonumber\\
   &&+\,F_2^{\mu}(P,K)\rho_{\mu\nu}(k_0,k)
   \tilde{\alpha}(p_0',\vec{p}-\vec{k})F_1^{\nu}(P,K)\nonumber\\
   &&+\,F_2^{\mu}(P,K)\rho_{\mu\nu}(k_0,k)
   \tilde{\beta}(p_0',\vec{p}-\vec{k})F_2^{\nu}(P,K)\nonumber\\
   &&+\left. G^{\mu\nu}(P,K)
   \rho_{\mu\nu}(k_0,k)\right\}_{p_0=\omega_p}\, ,
   \label{eq:prodspec}
\ee
where $\alpha_g=g^2/(4\pi)$ and the functions $\rho_{\mu\nu},\tilde{\rho},
\tilde{\alpha},\tilde{\beta},F_1^{\mu},F_2^{\mu}$ and $G^{\mu\nu}$ are defined 
in the appendix. Eq.~(\ref{eq:prodspec}) develops infrared divergences 
associated with the exchange of unscreened long wavelength magnetic gauge 
bosons. Nevertheless, it has been shown that the leading divergences can be 
resummed by a generalization of the Bloch-Nordsieck model at finite 
temperature~\cite{Blaizot}. Eq.~(\ref{eq:prodspec}) is valid for any $SU(N)$ 
gauge group, provided that the thermal masses include the corresponding group 
factors, and the generalization to the case with finite chemical potentials 
is also straightforward~\cite{LeBellac2}.

When computing the on-shell matrix element 
$\bar{u}(P){\mbox I}{\mbox m}^\star\Sigma_0(\omega_p,p)u(P)$ 
using the medium spinors satisfying the normalization condition in 
Eq.~(\ref{eq:proj}), the right-hand side of Eq.~(\ref{eq:prodspec}) can be 
identified as a sum of rates for processes involving thermalized fermions and 
gauge bosons. To this end, we note that the transverse
and longitudinal projection operators correspond to the products 
of transverse (summed over components) and of longitudinal photon 
polarizations, respectively, and that 
\be
   (\gamma^0\mp\vec{\gamma}\cdot\hat{p}) = \frac{1}{\omega_p^{\pm}Z_p^{\pm}}
   \sum_s u_s^{\pm}(\omega_p^{\pm},\vec{p})\bar{u}_s^{\pm}
   (\omega_p^{\pm},\vec{p})\, ,
   \label{spinsum} 
\ee
where the sum is over helicities of the (medium) fermion modes 
and the $\pm$ on the r.h.s.\ of Eq.~(\ref{spinsum}) refer to modes with either 
positive or negative helicity over chirality ratio. The functions $F_1^{\mu}$, 
$F_2^{\mu}$ and $G^{\mu\nu}$ are effective vertices and the terms containing 
$\tilde{\alpha}$ or $\tilde{\beta}$ represent products of amplitudes with an 
intermediate fermion propagator. Therefore, we write 
\be 
   \bar{u}(P){\mbox I}{\mbox m}^\star\Sigma_0(\omega_p,p)u(P) 
   = -\omega_p\Gamma\, ,
   \label{eq:gamdef}
\ee
where $\Gamma$ is the total reaction rate, which is the sum of decay and 
creation rates. Detailed balance ensures that the rate for the decay 
processes $\Gamma_D$, and the rate for the creation processes $\Gamma_I$, 
are related by $\Gamma_D=e^{\omega_p/T}\Gamma_I$. Eq.~(\ref{eq:gamdef})
was first computed, for the case $p=0$, in Refs.~\cite{Braaten2,Kobes}.  
Comparing Eqs.~(\ref{eq:newesp2}) and (\ref{eq:gamdef}) we finally obtain the 
relation
\be
   \gamma=\Gamma\, .
   \label{eq:finrel}
\ee
Three different kinds of products make up the total rate: the pole-pole, 
pole-cut and cut-cut contributions~\cite{Braaten3}. The pole-pole terms 
represent processes involving two quasiparticles, a fermionic excitation with 
either positive or negative helicity over chirality ratio and a gauge boson 
excitation, either transverse or longitudinal, in addition to the original 
fermion mode. The pole-cut terms represent either scattering or annihilation 
processes involving the exchange of a fermion or a gauge boson mode. Finally 
the cut-cut terms represent the same scattering or annihilation processes as 
the pole-cut terms together with the radiation of a soft gauge boson. 
The physical origin of the two latter kinds of processes is Landau 
damping~\cite{Landau}. For specific values of the energy and momentum of the 
original fermion mode, some of these processes will be kinematically 
forbidden. It is easy to check that by restricting our attention 
to the pole-pole contributions in the first term of Eq.~(\ref{eq:prodspec}), 
one obtains the analogue to Eq.~(35) in Ref.~\cite{D'Olivo}. We emphasize that,
although our expression for $\Gamma$ in Eq.~(\ref{eq:gamdef}) is formally the 
same as the one used in Ref.~\cite{Weldon1}, the original fermion mode is 
represented by the spinor $u(P)$ corresponding to a quasiparticle propagating 
in the medium, {\it and not}\/ by a free-particle spinor, from which it 
differs by the normalization.

\section{Hard axion production in a hot QED plasma}

As an application, consider the computation of the production rate 
of hard axions in a hot QED plasma through the Compton-like scattering 
diagram depicted in Fig.~2. This process is free from infrared 
divergences and thus it better illustrates the necessity of including 
the correct fermion wave function normalization in the calculation.

\begin{figure}[b]
\vspace{.2in}
\centerline{\epsfig{file=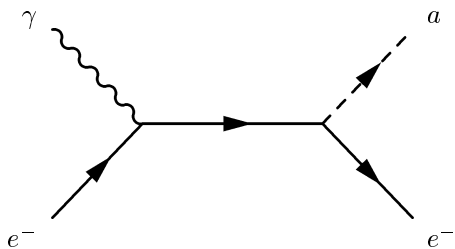}}
\vspace{.2in}
\caption{ Axion photoproduction by Compton-like scattering. }
\end{figure}

The tree level calculation involves an integral over the three momentum
transferred $p'$ of the virtual electron. Since in the soft $p'$ region 
HTL corrections to the electron propagator are not suppressed by powers 
of $e$, we need to use the effective electron propagator. Also, notice 
that since the axion momentum $k$ is hard, then, for soft $p'$, the 
momentum of the incoming electron $p$ must also be hard. The 
contribution to the hard-axion production rate, $\Gamma (k)$, from 
the soft $p'$ region is most conveniently found from the imaginary 
part of the axion self energy depicted in Fig.~3. The heavy dot in 
the loop electron line represents the effective propagator.

\begin{figure}[b]
\vspace{.2in}
\centerline{\epsfig{file=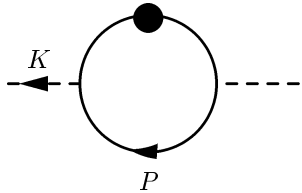}}
\vspace{.2in}
\caption{ Lowest order hard axion self-energy. The heavy dot in one of
          the internal electron lines represents the HTL effective electron 
          propagator. }
\end{figure}

 In Euclidean space, the expression for the axion self-energy 
is written as
\be
   \Pi (k) = \lambda^2 T\sum_n\int\frac{d^3p'}{(2\pi)^3}
   {\mbox T}{\mbox r}[K\!\dia\!\!\gamma_5 S(P')S_H(P'+K)K\!
   \dia\!\!\gamma_5]\, , 
   \label{axself}
\ee
where $\lambda$ is the effective axion-electron coupling with units 
of inverse energy and $S_H$ is the hard eletron propagator. 
Given that at least one of the momenta flowing into 
the vertices is hard, there is no need to consider HTL corrections to 
the vertices. $\Gamma (k)$ is related to $\Pi (k)$ by
\be
   \Gamma (k) = \frac{1}{k_0}\frac{ {\mbox I}{\mbox m}\Pi (k) }
   {(1-e^{k_0/T})}\, ,
   \label{ew:gampi}
\ee
where the factor $(1-e^{k_0/T})^{-1}$ selects the processes in which 
axions are produced. We now use the analog of Eq.(\ref{eq:disc}) for the 
fermion-antifermion case to compute the imaginary part of the axion 
self-energy. The terms involved are
\be
   {\mbox I}{\mbox m} T\sum_n S(P')S_H(P'+K)&=&\pi (1-e^{k_0/T})
   \int_{\infty}^{\infty}\frac{dp_0'}{2\pi}
   \int_{\infty}^{\infty}\frac{dp_0}{2\pi}\nonumber\\
   &\times &\tilde{f}(p_0')\tilde{f}(-p_0)\delta(k_0+p_0'-p_0)
   \tilde{\rho}(p_0')\tilde{\rho}(p_0)\, ,
   \label{eq:disc2}
\ee
where in the right-hand side we have performed the analytical 
continuation to Minkowsky space. Notice that\cite{LeBellac2} since 
$p$, $p_0$ are hard and $p_0^2-p^2>0$
\be
   \rho_{\pm}(p_0,p)=2\pi Z_+(p)\delta(p_0\mp \omega_+(p))\, ,
   \label{eq:specvac}
\ee
since for hard momentum, the modes with negative helicity over chirality
ratio decouple. Keeping only the contribution from positive energy 
electrons, the hard axion production rate in the soft $p'$ region is 
written as 
\be
   \Gamma_{soft}(k)&=&\frac{\lambda^2\pi}{k}\int\frac{d^3p'}{(2\pi )^3}
   \int_{\infty}^{\infty}\frac{dp_0'}{2\pi}
   \int_{\infty}^{\infty}\frac{dp_0}{2\pi}
   \tilde{f}(p_0')\tilde{f}(-p_0)\delta(k_0+p_0'-p_0)\nonumber\\
   &\times &2\pi Z_+(p)\delta (p_0-\omega_+(p))
   {\mbox T}{\mbox r}[K\!\dia\!\tilde{\rho}(p_0')K\!\dia\! (\gamma_0-
   \vec{\gamma}\cdot\hat{p})]\, .
   \label{eq:rasoft}
\ee
After integration over the direction of $\vec{p}\,'$ in 
Eq.~(\ref{eq:rasoft}), we obtain two kinematical restrictions for 
$p_0'$ and $p'$, namely ${p'}^2-{p'_0}^2 > 0$ and $2k > p'-p_0'$, 
therefore, after some simplifications
\be
   \Gamma_{soft}(k)&=&\frac{\lambda^2}{8k\pi^4}\int_0^{q^*}
   {p'}^2dp'\int_{-p'}^{p'}dp_0'\tilde{f}(p_0')\tilde{f}(-p_0'-k)
   \frac{(p_0'+k)}{2p'k}\theta (2k+p_0'-p')Z_+(p)\nonumber\\
   &\times &\left\{ \beta_+(p_0',p'){\mbox T}{\mbox r}
   [K\!\dia\!(\gamma_0-\vec{\gamma}\cdot\hat{p}')
   K\!\dia\!(\gamma_0-\vec{\gamma}\cdot\hat{p})] \right.\nonumber\\
   &+&\left. 
   \beta_-(p_0',p'){\mbox T}{\mbox r}
   [K\!\dia\!(\gamma_0+\vec{\gamma}\cdot\hat{p}')
   K\!\dia\!(\gamma_0-\vec{\gamma}\cdot\hat{p})]\right\}\, ,
   \label{eq:intrasoft}
\ee
where we have introduced the intermediate scale $q^*$ with 
$eT\ll q^* \ll T$\cite{Yuan}. Carring out the calculation of the traces 
and using that for hard $k$
\be
   \tilde{f}(-p_0'-k)=\frac{1}{e^{-(k_0+p_0')/T}+1}
   \approx\frac{1}{e^{-k/T}+1}\approx 1\, ,
   \label{eq:aprx}
\ee
we get
\be
   \Gamma_{soft}(k)&=&\frac{\lambda^2}{8k\pi^4}\int_0^{q^*}dp'
   \int_{-p'}^{p'}dp_0'\tilde{f}(p_0')Z_+(p_0'+k)\theta (2k+p_0'-p')
   ({p'}^2-{p_0'}^2)\nonumber\\
   &\times &\left\{(p'-p_0')(2k+p_0'+p')\beta_+(p_0',p') +
   (p'+p_0')(2k+p_0'-p')\beta_-(p_0',p')\right\}\, .
   \label{eq:gamaxi}
\ee
We notice that Eq.~(\ref{eq:gamaxi}) involves a factor $Z_+(p_0'+k)$ 
which arises from cutting the electron line with hard momentum in the 
loop diagram of Fig.~3. This factor signals that the electron at hand 
corresponds to an excitation in the medium. 

We now ask whether the same expression for the hard axion production 
rate can be recovered from considering the self-energy of this hard 
electron in the medium. As we will show, the expression thus obtained 
will coincide with Eq.~(\ref{eq:gamaxi}) provided that we include the 
proper normalization of the medium electron.

\begin{figure}[b]
\vspace{.2in}
\centerline{\epsfig{file=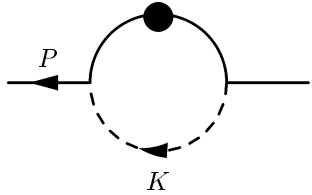}}
\vspace{.2in}
\caption{ Lowest order hard electron self-energy. The heavy dot in
          the internal electron line represents the HTL effective electron 
          propagator. }
\end{figure}

Consider the diagram in Fig.~4 representing the self-energy 
of a hard electron in the medium. The dotted line represents the axion. 
As before, in the soft $p'$ region, HTL corrections to the electron 
propagator are not suppressed by powers of $e$ and have to be resummed 
into the effective electron propagator represented by the line with 
the heavy dot. In Eucledian space, the expression for the electron 
self-energy is
\be
   \Sigma (P)=-\lambda^2 T \sum_n \int\frac{d^3p'}{(2\pi)^3}
   (P\!\dia\!+ P\!\dia\!')\gamma_5 G(P+P')S(P')(P\!\dia\!
   + P\!\dia\!')\gamma_5\, ,
   \label{eq:sigmele}
\ee
where $G(P)$ is the bare axion propagator. Production of axions is 
equivalent to destruction of the original hard electrons. The rate of 
destruction of these electrons is given, according to 
Eqs.~(\ref{eq:some2}) and (\ref{eq:finrel}), by
\be
   \Gamma (p)=-Z_+(p)\frac{ {\mbox T}{\mbox r}
   [(\gamma_0 - \vec{\gamma}\cdot\hat{p}){\mbox I}{\mbox m}\Sigma(P)]}
   {(1+e^{-p_0/T})}\, ,
   \label{eq:gamsi}
\ee
where the factor $(1+e^{-p_0/T})^{-1}$ selects the processes where 
the hard electron disappears from the medium and we have also summed over 
helicity states of the incoming electron. Notice that in 
Eq.~(\ref{eq:gamsi}) we have included explicitly the factor $Z_+(p)$, 
as corresponds to a rate that involves medium electrons. 
Using Eq.~(\ref{eq:disc}) we have
\be
   {\mbox I}{\mbox m} T\sum_n G(P+P')S(P')&=&
   \pi (e^{p_0/T}+1)\int_{\infty}^{\infty}\frac{dk_0}{2\pi}
   \int_{\infty}^{\infty}\frac{dp_0'}{2\pi}
   f(k_0)\tilde{f}(-p_0')\nonumber\\
   &\times &\delta (p_0-k_0+p_0')
   \tilde{\rho}(p_0')2\pi\delta [k_0^2-(\vec{p}+\vec{p}')^2]\, ,
   \label{eq:improd}
\ee
where we used that $2{\mbox I}{\mbox m}G(P+P')=2\pi\delta [(P+P')^2]$
and have also performed the analytical continuation to Minkowsky space.
From Eqs.~(\ref{eq:gamsi}) and (\ref{eq:improd}) we get after carrying 
out the angular integration
\be
   \Gamma_{soft} (p)&=&\frac{\lambda^2}{16\pi^2}
   \int_0^{q^*}dp'p'\int_{-p'}^{p'}dp_0'\theta (2p+p_0'-p')
   \frac{Z_+(p)}{p}e^{p_0/T}f(p_0+p_0')\tilde{f}(-p_0')\nonumber\\
   &\times &
   {\mbox T}{\mbox r}[(\gamma_0 - \vec{\gamma}\cdot\hat{p})
   (P\!\dia\! + P\!\dia\!')\tilde{\rho}(p_0')(P\!\dia\! + 
   P\!\dia\!')]\, ,
   \label{eq:gamintp}
\ee
where the integration over $p'$ is up to the intermediate scale $q^*$.
Notice that since $p_0$ is hard and $p_0'$ is soft, 
\be
   e^{p_0/T}f(p_0+p_0')=\frac{e^{p_0/T}}{e^{(p_0+p_0')/T}-1}
   \approx \frac{e^{p_0/T}}{e^{p_0/T}}=1\, .
   \label{eq:ide}
\ee
Thus, after computing the traces involved, we get
\be
   \Gamma_{soft} (p)&=&\frac{\lambda^2}{16\pi^4}
   \int_0^{q^*}dp'p'\int_{-p'}^{p'}dp_0'\theta (2p+p_0'-p')
   \tilde{f}(-p_0')\frac{Z_+(p)({p'}^2-{p_0'}^2)}{p^2}\nonumber\\
   &\times &
   \left\{\beta_+(p_0',p')(p'-p_0')(2p-p'+p_0') +
   \beta_-(p_0',p')(p'+p_0')(2p+p'+p_0')\right\}\, .
   \label{eq:gamintdosp}
\ee
In order to transform the rate of electron loss into the rate of axion
production, we use that $p=k-p_0'$, as required by kinematics,
and that $k$ is hard. Therefore,
\be
   \frac{1}{p^2}&=&\frac{1}{(k-p_0')^2}\approx\frac{1}{k^2}\nonumber\\
   Z_+(p)&=&Z_+(k-p_0')\nonumber\\
   (2p+p_0'\mp p')&=&(2k-p_0'\mp p')\, .
   \label{eq:consmom}
\ee
We now introduce unity in the form 
\be
   1=\int d^4k\delta(K-P-P')\, ,
   \label{eq:unity}
\ee
to write
\be
   \Gamma_{soft}(p)=\int d^4k\delta(K-P-P')\Gamma_{soft}(k)
   \label{eq:deliden}
\ee
where $\Gamma_{soft}(k)$ is given by
\be
   \Gamma_{soft}(k)&=&\frac{\lambda^2}{8k^2\pi^4}
   \int_0^{q^*}dp'\int_{-p'}^{p'}dp_0'\tilde{f}(-p_0')
   Z_+(k-p_0')\theta (2k-p_0'-p')({p'}^2-{p_0'}^2)\nonumber\\
   &\times &
   \left\{(p'-p_0')(2k-p_0'-p')\beta_+(p_0',p') +
   (p'+p_0')(2k-p_0'+p')\beta_-(p_0',p')\right\}\, .
   \label{eq:samegam}
\ee
Eq.~(\ref{eq:samegam}) coincides with Eq.~(\ref{eq:gamaxi}) 
as can be checked by means of the substitution $p_0'\rightarrow -p_0'$ 
for which $\beta_+\leftrightarrow\beta_-$.

It is now straightforward to compute the explicit expression for 
$\Gamma_{soft}(k)$. We use that in the soft $p'$, $p_0'$ region
$\tilde{f}(p_0')\approx 1/2$ and that $k\gg p_0$, $p_0'$, therefore
\be
   \Gamma_{soft}(k)&=&\frac{\lambda^2 Z_+(k)}{8k\pi^4}
   \int_0^{q^*}dp'\int_{-p'}^{p'}dp_0'({p'}^2-{p_0'}^2)\nonumber\\
   &\times &
   \left\{(p'-p_0')\beta_+(p_0',p') + (p'+p_0')\beta_-(p_0',p')
   \right\}\, .
   \label{eq:gamexpsof}
\ee
The integral is computed by using the sum rules satisfied by 
$\beta_{\pm}$~\cite{LeBellac2}. The result is
\be
   \Gamma_{soft}(k)&=&\frac{\lambda^2 Z_+(k)}{2k\pi}
   m_f^4\left\{\frac{q^{*2}}{m_f^2}
   -\int_0^\infty dx\left[\left( x^3-y_+(x)x^2-y_+^2(x)x
   +y_+^3(x)\right)Z_+(x)\right.\right.\nonumber\\
   &+&\left.\left.\left( x^3+y_-(x)x^2-y_-^2(x)x
   -y_-^3(x)\right)Z_-(x)\right]\right\}\, ,
   \label{eq:intgamsof}
\ee
where we have defined the dimensionless function 
$y_\pm(x)=\omega_\pm(x)/m_f$ and have extended the limit of integration 
in the second term to $\infty$. The remaining integral in 
Eq.~(\ref{eq:intgamsof}) is also a dimensionless
quantity that can be computed numerically and its value is 
approximately $1.3$.

We now proceed to compute the contribution to the hard axion
production rate from the hard momentum transfer region. 
For this case, the tree level electron propagator
must be used. Notice that the tree level propagator can be recovered
from the effective propagator by simply neglecting $m_f^2$ in the 
denominators of Eqs.~(\ref{eq:al}). In this manner, the approximate 
spectral densities become
\be
   \beta_\pm (p_0',p')\approx\frac{\pi^3m_f^2/{p'}^2}{(p'\mp p_0')}\, .
   \label{eq:treefer}
\ee
Thus, the contribution to the axion production rate from the hard
$p'$ region is
\be
   \Gamma_{hard}(k)=\frac{\lambda^2 Z_+(k)}{2k\pi}m_f^2
   \int_{q^*}^{\infty}dp'\int_{-p'}^{p'}dp_0'\tilde{f}(p_0')
   \frac{({p'}^2-{p_0'}^2)}{{p'}^2}\theta (2k+p_0'-p')\, .
   \label{eq:gamhar}
\ee
It is convenient to further decompose the hard region $(p'>q^*)$
into a $(I)$ low $(p_0'<q^*)$ and a $(II)$ high $(p_0'>q^*)$ 
frequency regions. In the low frequency region, we can still use
the approximation $\tilde{f}(p_0')\approx 1/2$. The remaining 
integrals are readily performed and the result is
\be
   \Gamma_{hard}^{(I)}(k)=\frac{\lambda^2 Z_+(k)}{2k\pi}m_f^2\,
   ({-q^*}^2)\, .
   \label{eq:gamI}
\ee
In the high frequency region, $\tilde{f}$ provides the cutoff
for the integrals. Recalling that $q^*\ll T$ we get
\be
   \Gamma_{hard}^{(II)}(k)=\frac{\lambda^2 Z_+(k)}{2k\pi}m_f^2
   \left\{ 2k^2 + 2kT\ln (1+e^{-k/T}) \right\}\, .
   \label{eq:gamII}
\ee
Adding Eqs.~(\ref{eq:intgamsof}), (\ref{eq:gamI}) and 
(\ref{eq:gamII}), the dependence on the intermediate scale
$q^*$ cancels, as it should, and the final expression for
the hard axion production rate is written as
\be
   \Gamma (k)=\frac{\lambda^2 Z_+(k)}{k\pi}m_f^4
   \left\{\frac{k^2}{m_f^2} + \frac{kT}{m_f^2}\ln (1+e^{-k/T})
   - 0.65\right\}\, ,
   \label{eq:finalgam}
\ee 
which is valid for $k\sim T$. Inclusion of a possible electron chemical 
potential is straightforward. Equation~(\ref{eq:finalgam}) represents the 
leading contribution to the photoproduction rate of axions in relativistic
plasmas~\cite{Hatsuda} at large $k$. Though this process is known to be  
subdominant as compared to nucleon bremsstrahlung of axions in this kind of 
plasmas~\cite{Raffelt}, a complete leading order treatment such as that of
Eq.~(\ref{eq:finalgam}) could have significant consequences for astrophysical
processes in strongly magnetized relativistic plasmas~\cite{Borisov}.

We should enphasize that in order for the rate obtained from the
axion self-energy to formally coincide with the rate obtained from the
electron self-energy, the latter has to be weighed with the factor
$Z_+$ as corresponds to a medium electron mode. 

Finally, we note that for hard $k$\cite{LeBellac2}
\be
   Z_+(k)\approx 1+ \frac{m_f^2}{2k^2}\left(
   1-\ln\left(\frac{2k^2}{m_f^2}\right)\right)\, ,
   \label{eq:zetapr}
\ee
thus, the error involved when ignoring the factor $Z_+$ is
small. This is in contrast to the situation when $k$ is soft for
which
\be
   Z_\pm(k)\approx \frac{1}{2}\pm \frac{k}{3m_f}\, .
   \label{eq:zetapr2}
\ee

\section{Discussion}

In this paper, we have argued, in the realm of hot gauge theories, the equality 
of the damping rate of a fermion mode propagating in a medium and its total 
reaction rate, to leading order, provided the latter is computed using the
wave function of the mode in the medium. This result is 
based on the explicit gauge independence of the on-shell matrix element of the
absorptive part of the fermion self-energy, according to the effective 
perturbative expansion of Braaten and Pisarski. It also unifies two apparently 
distinct concepts, both related to ${\mbox I}{\mbox m}\,^\star\Sigma_0$. In 
addition, Eq.~(\ref{eq:finrel}) also sheds light on the distinct interpretation
of a quantity like $\Gamma$ in the work of Weldon~\cite{Weldon1} and in 
ours. In effect, according to Eq.~(\ref{eq:proj}), the spinors $u(P)$ are
normalized to $Z_p$, related to the probability of finding a fermion mode
with energy $\omega_p$ and momentum $\vec{p}$ pertaining to the medium. On the 
other hand, if the spinor $u(P)$ was taken as free, it would represent a
mode in vacuum. Thus, if we take spinors from the medium, $\Gamma$ can be 
interpreted as the rate to approach equilibrium for an excitation in the
medium --- slightly out of equilibrium --- which, according to 
Eq.~(\ref{eq:finrel}), is equivalent to the rate at which the excitation is 
damped. On the other hand, if the spinors are from vacuum, the corresponding 
reaction rate can be interpreted as the rate to approach equilibrium for a test
particle, taken from the vacuum and dropped into the medium.

We have used the equivalence between damping and reaction rates of 
fermions in a medium to compute the hard axion production rate in a hot 
QED plasma by Compton-like scattering processes. We have shown that the 
rate obtained from the axion self-energy coincides with the rate 
obtained from the electron self-energy only when the proper medium
wave function is used in this latter computation.

Another interesting point, that we do not touch upon here, has 
to do with whether Eq.~(\ref{eq:finrel}) is a general result,
independent of perturbation theory, which is related to the way the 
thermalization of gauge degrees of freedom is treated~\cite{D'Olivo2}. 
The fact that at least to lowest, but complete order, the expression for the 
fermion damping rate in hot gauge theories is identical in the Coulomb or in 
covariant gauges, hints towards the validity of the result to all orders.

\section*{Acknowledgments}

We would like to thank J. Nieves for useful and stimulating discussions.
Support for this work has been received in part by CONACyT-M\'exico under
grants Nos. I27212 and 3298P-E9608.

\section*{Appendix}

Here, we list all of the functions introduced in section IV. Capital letters 
are used throughout to denote momentum four-vectors, all of them in Minkowski
space and $\hat{Q}^{\mu}=(-1,\hat{q})$. We start with the functions 
$F_1^{\mu}$, $F_2^{\mu}$ and $G^{\mu\nu}$, related to the explicit expressions 
for the HTL effective vertices 
$\Gamma_{\mu}$ and $\Gamma_{\mu\nu}$~\cite{LeBellac2}. 
\be
   F_1^{\mu}(P,K)&=&-m_f^2\int\frac{d\Omega}{4\pi}
   \frac{\hat{Q}^{\mu}\hat{Q}\dia}{[P\cdot\hat{Q}]}
   {\mathcal P}\left(\frac{1}{[(P-K)\cdot\hat{Q}]}\right)\, ,
   \nonumber\\
   F_2^{\mu}(P,K)&=&-m_f^2\int\frac{d\Omega}{4\pi}
   \frac{\hat{Q}^{\mu}\hat{Q}\dia}{[P\cdot\hat{Q}]}
   \delta((P-K)\cdot\hat{Q})\,\nonumber\\
   G^{\mu\nu}(P,K)&=&-m_f^2\int d\Omega
   \frac{\hat{Q}^{\mu}\hat{Q}^{\nu}\hat{Q}\dia}{[P\cdot\hat{Q}]^2}
   \delta((P-K)\cdot\hat{Q})\, .
   \label{eq:f1}
\ee
${\mathcal P}$ denotes the Cauchy principal value~\cite{Wong}. These
functions depend on the quantity 
$\Delta=P^2(P-K)^2-(P\cdot(P-K))^2$ and vanish for $\Delta > 0$.

Next, let us define the spectral densities, $\rho_{\mu\nu}$, $\tilde{\rho}$, 
$\tilde{\alpha}$ and $\tilde{\beta}$,
\be
   \rho_{\mu\nu}(k_0,k)&=&\frac{k^2}{k_0^2-k^2}\rho_L(k_0,k)P^L_{\mu\nu}+
   \rho_T(k_0,k)P^T_{\mu\nu}\nonumber\\
   \tilde{\rho}(p_0,\vec{p})
   &=&\rho_+(p_0,p)(\gamma_0-\vec{\gamma}\cdot\hat{p}) +
   \rho_-(p_0,p)(\gamma_0+\vec{\gamma}\cdot\hat{p})\nonumber\\
   \tilde{\alpha}(p_0,\vec{p})
   &=&\alpha_+(p_0,p)(\gamma_0-\vec{\gamma}\cdot\hat{p}) +
   \alpha_-(p_0,p)(\gamma_0+\vec{\gamma}\cdot\hat{p})\nonumber\\
   \tilde{\beta}(p_0,\vec{p})
   &=&\beta_+(p_0,p)(\gamma_0-\vec{\gamma}\cdot\hat{p}) +
   \beta_-(p_0,p)(\gamma_0+\vec{\gamma}\cdot\hat{p})\, ,
   \label{eq:esdenex}
\ee
where $P^L$ and $P^T$ are the longitudinal and transverse projection operators 
in four dimensions. The quantities $\rho_{L,T}$, $\rho_{\pm}$ are the standard 
HTL spectral densities corresponding to the gauge boson and fermion propagator,
respectively~\cite{LeBellac2}. We have also defined
\be
   \alpha_{\pm}(p_0,p)=\frac{(2\pi)\left[p_0\mp p-\frac{m_f^2}{p}
   \left( \frac{(p\mp p_0)}{2p}\ln \left|\frac{p_0+p}{p_0-p}\right|\pm 1
   \right)\right]\theta(p^2-p_0^2)}
   {\left[p_0\mp p-\frac{m_f^2}{p}\left( \frac{(p\mp p_0)}{2p}\ln 
   \left|\frac{p_0+p}{p_0-p}\right|\pm 1\right)\right]^2+
   \left[\frac{\pi m_f^2}{2p}\left(1\mp \frac{p_0}{p}\right)\right]^2}\, ,
   \label{eq:al0}
\ee
\be
   \beta_{\pm}(p_0,p)=-\frac{\pi^3\frac{m_f^2}{p}
   \left(\frac{p\mp p_0}{p}\right)\theta(p^2-p_0^2)}
   {\left[p_0\mp p-\frac{m_f^2}{p}\left( \frac{(p\mp p_0)}{2p}\ln
   \left|\frac{p_0+p}{p_0-p}\right|\pm 1\right)\right]^2+
   \left[\frac{\pi m_f^2}{2p}\left(1\mp \frac{p_0}{p}\right)\right]^2}\, .
   \label{eq:al}
\ee


\begin{thebibliography} {20}
\bibitem{explain} 
It is possible to use Eq.~(\ref{eq:disc}) since in Eq.~(\ref{eq:sigma}), the 
vertex $\Gamma_{\mu\nu}$ can be shown to depend on the four vectors
$P$ and $P-K$.
\end{thebibliography}
\end{document}